\documentclass[pra,twocolumn,superscriptaddress,showpacs,nobibnotes,notitlepage,nofootinbib]{revtex4-2}

\usepackage[utf8]{inputenc}
\usepackage[margin=1in]{geometry} 
\usepackage{amsmath,amsthm,amssymb,hyperref}
\hypersetup{
    colorlinks=true,
    linkcolor=blue,
    citecolor=magenta,
    filecolor=magenta,      
    urlcolor=blue,
    pdftitle={},
    pdfpagemode=FullScreen,
    }
\usepackage{adjustbox}
\usepackage{graphicx}
\usepackage{float}
\usepackage{xcolor}
\usepackage{soul}
\usepackage{qcircuit}

\usepackage{comment}
\usepackage{booktabs}

\usepackage{soul}
\usepackage{xcolor}
\sethlcolor{yellow}

\begin{document}

\title{Code switching revisited: Low-overhead magic state preparation using color codes}
\author{Lucas Daguerre}
\affiliation{Department of Physics and Astronomy, University of California, Davis, CA, 95616, USA}
\author{Isaac H. Kim}
\affiliation{Department of Computer Science, University of California, Davis, CA, 95616, USA }
\date{\today}

\begin{abstract}
We propose a protocol to prepare a high-fidelity magic state on a two-dimensional (2D) color code using a three-dimensional (3D) color code. Our method modifies the known code switching protocol with (i) a recently discovered transversal gate between the 2D and the 3D code and (ii) a judicious use of flag-based postselection. We numerically demonstrate that these modifications lead to a significant improvement in the fidelity of the magic state.  For instance, subjected to a uniform circuit-level noise of $10^{-3}$ (excluding idling noise), our code switching protocol yields a magic state encoded in the distance-$3$ 2D color code with a logical infidelity of $4.6\times 10^{-5}\pm 1.6 \times 10^{-5}$ (quantified by an error-corrected logical state tomography) with an $84\%$ of acceptance rate. Used in conjunction with a  postselection approach, extrapolation from a polynomial fit suggests a fidelity improvement to $5.1 \times 10^{-7}$ for the same code. Our protocol is aimed for architectures that allow nonlocal connectivity and should be readily implementable in near-term devices. Finally, we also present a simulation technique akin to an extended stabilizer simulator which effectively incorporates the non-Clifford $T$-gate, that permits to simulate the protocol without resorting to a resource intensive state-vector simulation.
\end{abstract}

\maketitle

\section{Introduction}
\label{sec:intro}

An essential step towards building a large-scale fault-tolerant quantum computer is quantum error correction~\cite{Shor1995}. In principle, the overhead associated with fault-tolerance is merely polylogarithmic in the size of the computation~\cite{Aharonov1997}, and can be even brought down to a constant~\cite{Gottesman2013}. However, the practical overhead needed for large-scale fault-tolerant quantum computation remains daunting.

It has been often stated in the literature that the most costly part of quantum computation is magic state distillation~\cite{Bravyi2005}. Traditional methods for magic state distillation relied on multi-level distillation of various protocols~\cite{Bravyi2005,Meier2012,Bravyi2012,Jones2013,Fowler2013,Jones2013a,Duclos2013,Duclos2015,Campbell2017,OGorman2017,Haah2018,Gidney2019}. Such analysis often led to an enormous overhead in preparing a magic state, motivating novel alternative methods. One of the important recent insights in reducing the overhead is that the size of the codes used at a lower level of distillation need not be as large~\cite{fowler2013bridgeloweroverheadquantum,Litinski2019}. 

Even more recently, there has been a line of works that aim to prepare a higher-fidelity magic state before the distillation steps. An initial work in this direction is due to Li, who pointed out that the quality of the magic state can be improved by a judicious postselection~\cite{Li2015}. Chamberland and Noh then used the flag-based methods~\cite{Yoder2017,Chao2018,Chao2018a} to prepare a magic state while detecting error with a small code~\cite{Chamberland2020}. While this work required a relatively challenging physical error rate of $10^{-4}$, recent improvements in this direction led to protocols that can achieve substantial reduction in error even at a realistic physical error rate of $10^{-3}$~\cite{Butt2024,Itogawa2024,Gidney2024}. These protocols all boast a rather modest requirement on the number of qubits, which make them attractive quantum error correction protocols to implement and test in the near term; see Ref.~\cite{Pogorelov2024experimental} for a recent experiment, for instance. These methods, together with recent efficient magic state distillation protocols~\cite{Litinski2019,Lee2024} are expected to further reduce the overhead in magic state distillation. In particular, Ref.~\cite{Gidney2024} reported a scheme that can reach an error rate as low as $2\times 10^{-9}$ (with the post-selection success probability of $1\%$) under the same physical error rate. This is a level of error that is sufficient to run some quantum algorithms that have low resource requirement, e.g., Ref.~\cite{Campbell2021}, obviating the need for magic state distillation entirely for such applications.

However, it should be noted that there are important quantum algorithms requiring a number of $T$-gates that is still larger than the distillation-free methods can tolerate. For instance, the number of Toffoli gates used in quantum chemistry applications such as FeMoCo ranges between $5.3\times 10^{9}$~\cite{Lee2021} and $1.0\times 10^{10}$~\cite{von2021}. Depending on the specifics of how the Toffoli gate is being used, each Toffoli can be converted to Clifford gates and between four and seven $T$ gates~\cite{Gidney2018halving}. Other applications in quantum chemistry tend to require even more $T$ gates~\cite{Kim2022,Delgado2022,Su2021}. Even with the most recent method~\cite{Gidney2024}, a completely distillation-free approach cannot yet reach an error rate required for these applications. Therefore, an outstanding question is whether it is possible to bring down the cost of $T$ gates even further, so that its cost is comparable to other Clifford gates in this regime. 

This paper introduces a different protocol that aims to achieve a similar goal. Unlike the recent protocols that aim to measure a transversal Clifford gate~\cite{Itogawa2024,Gidney2024}, our approach aims to exploit a code with a non-Clifford transversal gate. Such an approach was proposed in Refs.~\cite{Bombin2007,Bombin2016} and studied in detail in Ref.~\cite{Beverland2021}. However, in light of the recent works that improved the fidelity using postselection~\cite{Itogawa2024,Gidney2024}, it is natural to ask whether post-selection can also improve the fidelity of transversal-gate-based approaches.

To that end, we use the three-dimensional(3D) color code~\cite{Bombin2007}, and focus on its simplest instance, which is the $15$-qubit quantum Reed-Muller code (qRM). We will aim to use the transversal gate of the 3D color code, which is then converted to the 2D color code~\cite{Bombin2006,Bombin2016}. While a careful study has suggested that preparation of magic state using 3D color code using this approach is more costly than the multilevel distillation method~\cite{Beverland2021}, a more recent work exploited postselection and flag qubits~\cite{Yoder2017,Chao2018,Chao2018a} to improve the performance~\cite{Butt2024}. Our approach is similar in spirit, though we use the flag qubits in a different way and also uses a transversal gate between 3D color code and 2D color code (that was recently discovered in Ref.~\cite{sullivan2024}\footnote{We would like to thank the anonymous referees for pointing out this reference.}, which also appeared in Ref.~\cite{Heuben2024} during
the preparation of our manuscript.) Overall, our approach yields a high quality magic state on par with these prior results, which we discuss further in Sec.~\ref{sec:comparison}.

Although our protocol has a logical infidelity lower than what was reported in Refs.~\cite{Itogawa2024,Gidney2024} for similar setups (in terms of the code being used and the physical error rate), they are incomparable. References~\cite{Itogawa2024,Gidney2024} focused on quantum computing architectures equipped with nearest-neighbor gates on a two-dimensional grid, targeting superconducting qubit-based quantum computers~\cite{Kjaergaard2020superconducting}. Our approach uses gates that would require nonlocal connectivity. As such, our scheme, as it stands, cannot be used directly on these devices. 

We emphasize, however, that our protocol is aimed at other types of quantum computers, such as the ones based on photons~\cite{Knill2001scheme,Bartolucci2023fusion,Bourassa2021blueprintscalable}, ions~\cite{Circ1995,monroe2013scaling,bruzewicz2019trapped}, and neutral atoms~\cite{bluvstein2022quantum,Bluvstein2024}. Non-local gates can be accommodated in these architectures, making them promising candidates to implement our protocol.

Lastly, we remark that we introduce a simulation method for Clifford+$T$ circuits that may be of an independent interest. The main strength of our method is that it is directly applicable to existing stabilizer circuit simulators, such as Stim~\cite{Gidney2021}. The key ingredient relies on keeping track of the Pauli error frame associated to a logical stabilizer state. In this framework, a non-Clifford logical-$T$ gate can be mimicked by introducing appropriate Clifford errors, which can be simulated easily using the existing simulators~\cite{Gidney2021}. We provide more details in Sec.~\ref{sec:simulation} and Appendix~\ref{app:sim_details}.

The rest of this paper is organized as follows. In Sec.~\ref{sec:comparison}, we compare our protocol to the other related protocols in the literature. In Sec.~\ref{sec:review_color_codes}, we review aspects of color codes. In Sec.~\ref{sec:protocol}, we describe our protocol. In Sec.~\ref{sec:simulation} we describe our simulation method and present the result. We end with a discussion in Sec.~\ref{sec:discussion}. Finally, Appendix~\ref{app:sim_details} provides additional details about the simulation methods.

\section{Comparison with prior work}
\label{sec:comparison}

A comparison of the infidelity of magic states prepared in our protocol with some recent results in the literature can be found in Tables~\ref{tab:comparisson} and~\ref{tab:comparisson2}.

\begin{table*}[!ht]
\centering
\begin{tabular}{|c|c|c|c|c|}
\hline
\multicolumn{5}{|c|}{Uniform depolarizing noise}\\ \hline \hline
Method &  Infidelity & \begin{tabular}{@{}c@{}}Acceptance \\ rate\end{tabular} & Final code  & Distance\\ \hline \hline
\begin{tabular}{@{}c@{}}Our protocol \\ (error correction)\end{tabular} &$4.6\times 10^{-5}\pm 1.6 \times 10^{-5}$ & $84\%$ & Steane code & 3\\\hline
 Butt \textit{et al} ~\cite{Butt2024}&$ 10^{-4}$  & $94\%$ & Steane code & 3\\ 
\hline \hline
\begin{tabular}{@{}c@{}}Our protocol$^{(*)}$ \\ (postselection)\end{tabular} &$\approx 5.1 \times 10^{-7}$ & $\approx 84\%$ & Steane code & 3\\\hline
Gidney \textit{et al} ~\cite{Gidney2024} &$ 6\times 10^{-7}$ & $65\%$ & Steane code & 3\\ \hline
Itogawa \textit{et al} ~\cite{Itogawa2024} &$ 10^{-4}$ & $70\%$ & Surface code & 3\\
\hline
\end{tabular}
\caption{Infidelity of magic states $|\bar{T}\rangle$ under a uniform depolarizing error model with strength $p= 10^{-3}$, for some state-of-the-art protocols, prepared in the final stage on distance $d=3$ codes. Note that our error model does not include idling errors. The label ``error correction/postselection" implies that the infidelity of the magic states is quantified via state tomography, where the final state in the postprocessing stage is either error corrected or postselected if syndromes are trivial, respectively. The first two rows use error correction, while the last three rows use postselection. $^{(*)}$These results are \textit{extrapolated} from the data points of Figs.~\ref{fig:infidelity_plot} and \ref{fig:acceptance_plot}, following the polynomial fits of (\ref{eq:infidelity_fit}) and (\ref{eq:acceptance_fit}), therefore the approximate symbol $\approx$ is utilized.}
\label{tab:comparisson}
\end{table*}

\begin{table*}[!ht]
\centering
\begin{tabular}{|c|c|c|c|c|}
\hline
\multicolumn{5}{|c|}{Multiparameter depolarizing noise }\\ \hline \hline
Method &  Infidelity &  \begin{tabular}{@{}c@{}}Acceptance \\ rate\end{tabular} & Final code & Distance\\ \hline \hline
\begin{tabular}{@{}c@{}}Our protocol \\ (error correction)\end{tabular}  &$9\times 10^{-5} \pm 2\times 10^{-5}$& $71\%$ & Steane code & 3\\\hline
 Butt \textit{et al} ~\cite{Butt2024}&$ 1.6 \times 10^{-4}\pm 2\times 10^{-5}$  & $82\%$ & Steane code & 3\\ \hline
Postler \textit{et al} ~\cite{postler2022demonstration} &$ 10^{-4}\pm 3\times 10^{-5}$ & $85\%$ & Steane code & 3\\
\hline 
\end{tabular}
\caption{Infidelity of magic states $|\bar{T}\rangle$ ($|\bar{H}\rangle$ states for Ref.~\cite{postler2022demonstration}) under a multiparameter depolarizing error model with initialization and measurement error $p_i=p_m=10^{-3}$, single-qubit error $p_1=10^{-4}$, and two-qubit error $p_2=3\times 10^{-3}$, for some state-of-the-art protocols, prepared in the final stage on the Steane code $[[7,1,3]]$. Note that our error model does not include idling errors and that Ref.~\cite{postler2022demonstration} uses the single-qubit depolarizing noise model for initialization and measurement errors. The infidelity of these magic states are quantified via state tomography, where the final state in the post-processing stage is error corrected. The numerical values for the protocol \cite{postler2022demonstration} were obtained from Ref.~\cite{Butt2024}.}
\label{tab:comparisson2}
\end{table*}

A common feature of all these protocols is that they rely on some degree of postselection. However, there is also a difference in the overall approach. Our protocol (similar to Ref.~\cite{Butt2024}) uses a transversal non-Clifford gate, whereas Refs.~\cite{Itogawa2024,Gidney2024,postler2022demonstration} use the measurement of logical Clifford operators. 

For our work, an important caveat is that the logical error rate estimate for the magic state was based on an ``ungrown" patch of fixed distance $d=3$. In order to assess the quality of the magic state that will be used in practice, one needs to estimate the fidelity after converting this state to a magic state encoded in a larger-distance code~\cite{Gidney2024}. We instead focus on estimating the fidelity for the $d=3$ code; the error analysis related to the patch growth is left for future work. 

Nonetheless, we can compare our protocols to the other protocols in the literature that studied the quality of the magic state on an ungrown patch. There are two types of works in the literature, depending on how one quantifies the quality of the magic state. In the first approach, one performs a logical state tomography at the final step, by performing noiseless measurement on all the qubits followed by an error correction. In the second approach, one instead post-selects on observing a trivial syndrome. 

In the first approach, our numerical simulation reports an infidelity on par with other state-of-the-art results in the literature, using similar error models (no idling errors were included in our simulations); see Tables~\ref{tab:comparisson} and ~\ref{tab:comparisson2}. For the second approach, we were unable to directly simulate at a physical error rate used in the literature; the infidelity resulting from the second approach is too low. Instead, we fitted the logical error rate with a polynomial at higher physical error rate and extrapolated. Needless to say, the result obtained this way cannot be directly compared to the results obtained from exact diagonalization~\cite{Itogawa2024,Gidney2024}. Nonetheless, it can serve as a proxy to understand how well our method would work.

\section{Color codes}
\label{sec:review_color_codes}

\begin{figure}[ht]
    \centering
    \includegraphics[width=0.3\textwidth]{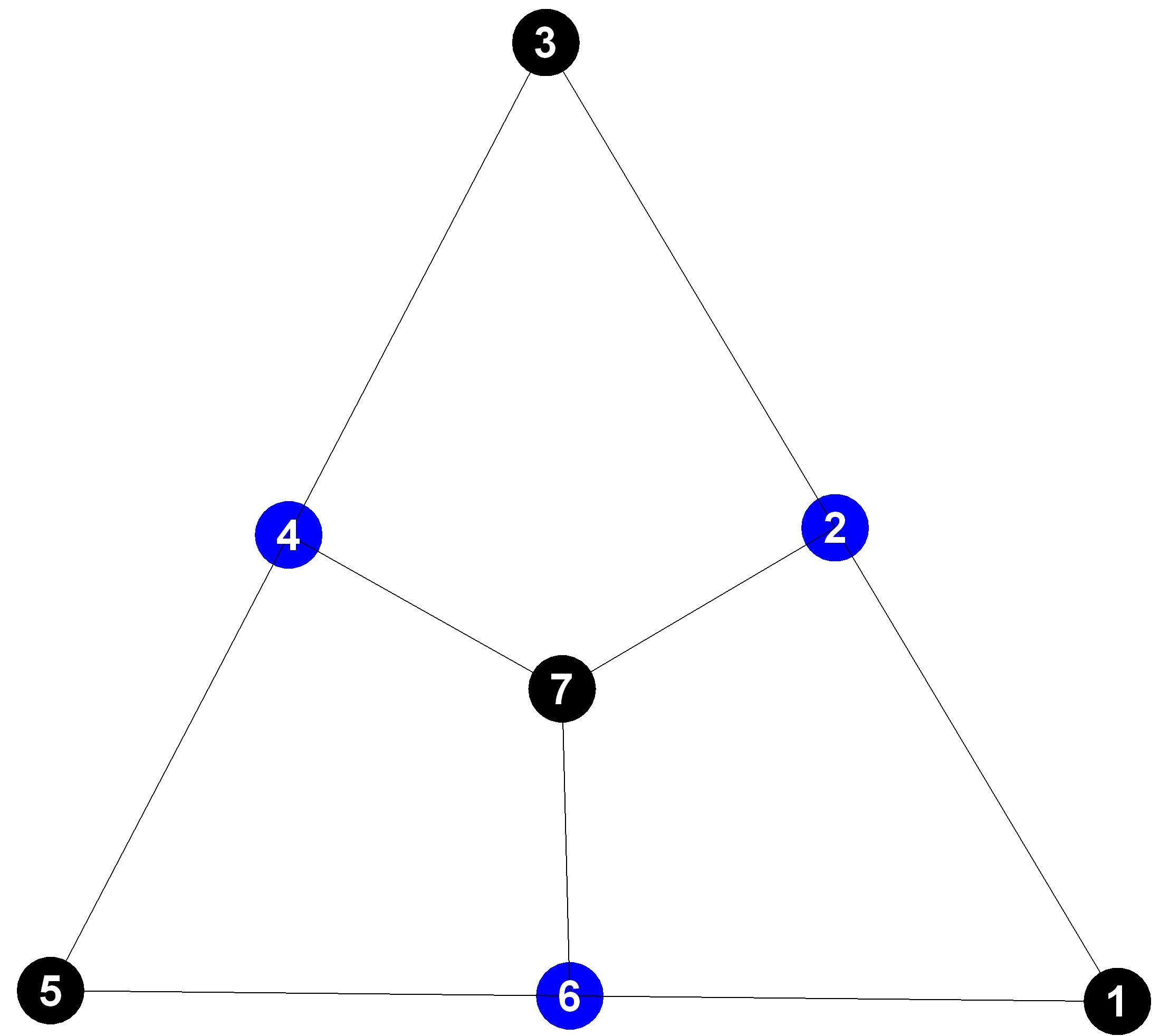}
    \caption{Pictorial representation of the $[[7,1,3]]$ Steane code including labels for the physical qubits.}
    \label{fig:Steane}
\end{figure}

\begin{figure}[!ht]
    \centering
    \includegraphics[width=0.3\textwidth]{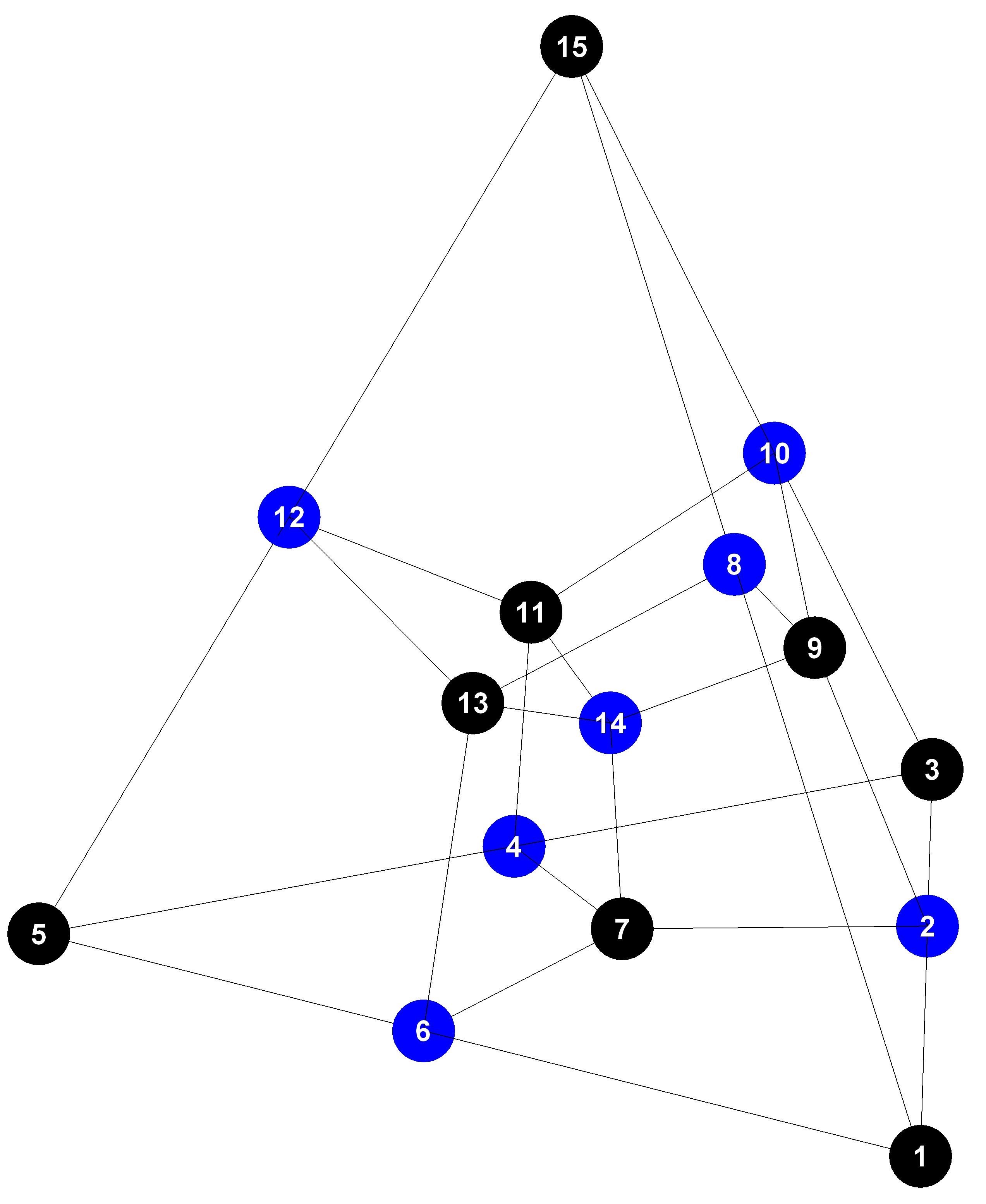}
    \caption{Pictorial representation of the $[[15,1,3]]$ qRM code including labels for the physical qubits.}
    \label{fig:qRM}
\end{figure}

In this section, we provide a brief review of the two- and three-dimensional color codes~\cite{Bombin2006,Bombin2007} that are relevant to this paper. This section will introduce a convention for the qubit labels we use, which shall be used throughout this paper; see Figs.~\ref{fig:Steane} and~\ref{fig:qRM}.

Most of the results in this section are well known, but the discussion about a transversal gate between the two codes is relatively new, having recently appeared in Refs.~\cite{sullivan2024} and~\cite{Heuben2024} (the later gate is equivalent to ours up to a transversal unitary).

The two-dimensional color code we consider is a CSS code with parameters $[[7,1,3]]$, also known as the Steane code~\cite{Steane1996}. The Steane code stabilizer group is generated by mutually commuting $Z$-type and $X$-type $2$-cells (henceforth referred to as plaquettes) stabilizers, described in Table~\ref{tab:stab_Steane}. 

\begin{table}[!ht]
\centering
\begin{tabular}{|cccc|}
\hline
\multicolumn{4}{|c|}{Steane: $Z(X)$-type stabilizers} \\ \hline
\multicolumn{1}{|c|}{$p_1$}      & \multicolumn{1}{|c|}{$Z_1Z_2Z_6Z_7$} & \multicolumn{1}{|c|}{$p_{4}$} & \multicolumn{1}{|c|}{$X_1X_2X_6X_7$} \\ \hline
\multicolumn{1}{|c|}{$p_2$}      & \multicolumn{1}{|c|}{$Z_2Z_3Z_4Z_7$} & \multicolumn{1}{|c|}{$p_{5}$} & \multicolumn{1}{|c|}{$X_2X_3X_4X_7$} \\ \hline
\multicolumn{1}{|c|}{$p_3$}      & \multicolumn{1}{|c|}{$Z_4Z_5Z_6Z_7$} & \multicolumn{1}{|c|}{$p_{6}$} & \multicolumn{1}{|c|}{$X_4X_5X_6X_7$} \\ \hline

\end{tabular}
\caption{$Z$-type and $X$-type plaquette stabilizer generators for the $[[7,1,3]]$ Steane code.}
\label{tab:stab_Steane}
\end{table}

The Steane code encodes a single logical qubit and has minimum weight-$3$ transversal logical $\bar{Z}$ and $\bar{X}$ operators
\begin{equation}
    \bar{Z}=Z_1Z_2Z_3\:,
\label{eq:logicalZSteane}
\end{equation}
\begin{equation}
    \bar{X}=X_1X_2X_3\:.
\label{eq:logicalXSteane}
\end{equation}
Given that the distance $d=3$, the Steane code is capable of correcting arbitrary weight-$1$ errors, and detecting arbitrary weight-$2$ errors. Unfortunately, the Steane code does not support any transversal logical non-Clifford gate like the $T$ gate. This is because the Steane code supports the full Clifford group transversally, however the no-go Eastin-Knill theorem \cite{eastin2009restrictions} states that no stabilizer quantum code exists that contains a transversal universal gate set. 

The three dimensional color code we consider is a CSS code with parameters $[[15,1,3]]$. It is the smallest color code in three dimensions that admits a transversal $T$-gate and represents a specific instance of a family of codes known as quantum Reed-Muller codes (qRM). 

The stabilizer group is generated by $Z$-type 2-cells (plaquettes) and $X$-type 3-cells. In total there are eighteen plaquettes (Table~\ref{tab:Z_stab_qRM}) and four 3-cells (Table~\ref{tab:X_stab_qRM}). Since the qRM encodes a single logical qubit and $15=1+10+4$, there are only ten independent $Z$-type stabilizers, meaning that there are eight independent constraints (Table~\ref{tab:qRM_constraints}). Those are given by multiplying plaquettes around a cylinder, yielding a total of three constraints per 3-cell. In a given 3-cell, two such constraints give rise to the third one, thus, two constraints per 3-cell give the net result of eight independent ones.

\begin{table}[!ht]
\centering
\begin{tabular}{|cccc|}
\hline
\multicolumn{4}{|c|}{qRM: $Z$-type stabilizers} \\ \hline
\multicolumn{1}{|c|}{$p_1$}      & \multicolumn{1}{|c|}{$Z_1Z_2Z_6Z_7$} & \multicolumn{1}{|c|}{$p_{10}$} & \multicolumn{1}{|c|}{$Z_6Z_7Z_{13}Z_{14}$} \\ \hline
\multicolumn{1}{|c|}{$p_2$}      & \multicolumn{1}{|c|}{$Z_2Z_3Z_4Z_7$} & \multicolumn{1}{|c|}{$p_{11}$} & \multicolumn{1}{|c|}{$Z_2Z_7Z_9Z_{14}$} \\ \hline
\multicolumn{1}{|c|}{$p_3$}      & \multicolumn{1}{|c|}{$Z_4Z_5Z_6Z_7$} & \multicolumn{1}{|c|}{$p_{12}$} & \multicolumn{1}{|c|}{$Z_4Z_7Z_{11}Z_{14}$} \\ \hline
\multicolumn{1}{|c|}{$p_4$}      & \multicolumn{1}{|c|}{$Z_1Z_6Z_8Z_{13}$} & \multicolumn{1}{|c|}{$p_{13}$} & \multicolumn{1}{|c|}{$Z_8Z_{12}Z_{13}Z_{15}$} \\ \hline
\multicolumn{1}{|c|}{$p_5$}      & \multicolumn{1}{|c|}{$Z_1Z_2Z_8Z_9$} & \multicolumn{1}{|c|}{$p_{14}$} & \multicolumn{1}{|c|}{$Z_8Z_9Z_{10}Z_{15}$} \\ \hline
\multicolumn{1}{|c|}{$p_6$}      & \multicolumn{1}{|c|}{$Z_2Z_3Z_9Z_{10}$} & \multicolumn{1}{|c|}{$p_{15}$} & \multicolumn{1}{|c|}{$Z_{10}Z_{11}Z_{12}Z_{15}$} \\ \hline
\multicolumn{1}{|c|}{$p_7$}      & \multicolumn{1}{|c|}{$Z_3Z_4Z_{10}Z_{11}$} & \multicolumn{1}{|c|}{$p_{16}$} & \multicolumn{1}{|c|}{$Z_8Z_9Z_{13}Z_{14}$} \\ \hline
\multicolumn{1}{|c|}{$p_8$}      & \multicolumn{1}{|c|}{$Z_4Z_5Z_{11}Z_{12}$} & \multicolumn{1}{|c|}{$p_{17}$} & \multicolumn{1}{|c|}{$Z_9Z_{10}Z_{11}Z_{14}$} \\ \hline
\multicolumn{1}{|c|}{$p_9$}      & \multicolumn{1}{|c|}{$Z_5Z_6Z_{12}Z_{13}$} & \multicolumn{1}{|c|}{$p_{18}$} & \multicolumn{1}{|c|}{$Z_{11}Z_{12}Z_{13}Z_{14}$} \\ \hline
\end{tabular}
\caption{$Z$-type plaquette stabilizers for the $[[15,1,3]]$ qRM code. Only ten of them are independent due to the constraints of Table~\ref{tab:qRM_constraints}. In the simulations, we decided to measure the set of ten independent stabilizers $\{p_1,p_2,p_3,p_7,p_8,p_9,p_{13},p_{16},p_{17},p_{18}\}$.}
\label{tab:Z_stab_qRM}
\end{table}

\begin{table}[!ht]
\centering
\begin{tabular}{|cc|}
\hline
\multicolumn{2}{|c|}{qRM: $X$-type stabilizers}        \\ \hline
\multicolumn{1}{|c|}{$c_1$} & $X_1X_2X_6X_7X_8X_9X_{13}X_{14}$ \\ \hline
\multicolumn{1}{|c|}{$c_2$}      &    $X_4X_5X_6X_7X_{11}X_{12}X_{13}X_{14}$   \\ \hline
\multicolumn{1}{|c|}{$c_3$}      &   $X_2X_3X_4X_7X_9X_{10}X_{11}X_{14}$    \\ \hline
\multicolumn{1}{|c|}{$c_4$}      &     $X_8X_9X_{10}X_{11}X_{12}X_{13}X_{14}X_{15}$  \\ \hline
\end{tabular}
\caption{$X$-type $3$-cell stabilizers for the $[[15,1,3]]$ qRM code.}
\label{tab:X_stab_qRM}
\end{table}

\begin{table}[!ht]
\centering
\begin{tabular}{|cccc|}
\hline
\multicolumn{4}{|c|}{qRM: Independent constraints}        \\ \hline
\multicolumn{1}{|c|}{$\Gamma_1$}  & \multicolumn{1}{|c|}{$p_4p_5p_{10}p_{11}$} & \multicolumn{1}{|c|}{$\Gamma_5$}     &     \multicolumn{1}{|c|}{$p_6p_7p_{11}p_{12}$}  \\ \hline
\multicolumn{1}{|c|}{$\Gamma_2$}      &    \multicolumn{1}{|c|}{$p_1p_5p_{10}p_{16}$}  & \multicolumn{1}{|c|}{$\Gamma_6$}      &     \multicolumn{1}{|c|}{$p_2p_6p_{12}p_{17}$}  \\ \hline
\multicolumn{1}{|c|}{$\Gamma_3$}      &   \multicolumn{1}{|c|}{$p_8 p_9 p_{10}p_{12}$}   & \multicolumn{1}{|c|}{$\Gamma_7$}      &     \multicolumn{1}{|c|}{$p_{13}p_{15}p_{16}p_{17}$}  \\ \hline
\multicolumn{1}{|c|}{$\Gamma_4$}      &     \multicolumn{1}{|c|}{$p_3p_8p_{10}p_{18}$}  & \multicolumn{1}{|c|}{$\Gamma_8$}      &     \multicolumn{1}{|c|}{$p_{14}p_{15}p_{16}p_{18}$}  \\ \hline
\end{tabular}
\caption{Independent set of constraints $\Gamma_i=I$ (identity) with $i=1,2,\ldots,8$ for the $[[15,1,3]]$ qRM code.}
\label{tab:qRM_constraints}
\end{table}

The minimum weight logical-$Z$ operator $\bar{Z}$ has weight-$3$, and a representative lies at the edge of the tetrahedron
\begin{equation}
    \bar{Z}=Z_1Z_2Z_3\:.
\label{eq:logicalZqRM}
\end{equation}
The minimum weight logical-$X$ operator $\bar{X}$ has weight-$7$, and a representative lies at the face of a boundary of the tetrahedron
\begin{equation}
    \bar{X}=X_1X_2X_3X_4X_5X_6X_7\:.
\label{eq:logicalXqRM}
\end{equation}
Hence, the distance $d=\text{min}\{d_x,d_z\}=3$, where $d_x=7$ is the distance for $X$-type logical operators, and $d_z=3$ is the distance for $Z$-type  logical operators. In addition, the $[[15,1,3]]$ qRM code is capable of correcting arbitrary weight-$1$ $Z$-errors (phase flip) and up to weight-$3$ $X$-errors (bit flip).

The qRM code also admits a transversal $T$ gate $\bar{T}$, consisting of single-qubit $T$/$T^{\dagger}$ gates applied to odd- and even labeled qubits, respectively,
\begin{equation}
\bar{T}=T_1T^{\dagger}_2T_3T^{\dagger}_4T_5T^{\dagger}_6T_7T^{\dagger}_8T_9T^{\dagger}_{10}T_{11}T^{\dagger}_{12}T_{13}T^{\dagger}_{14}T_{15}\:,
\label{eq:Tgate}
\end{equation}
where the single-qubit $T$ gate is defined as
\begin{equation}
    T_j=\begin{pmatrix}
    1 &0 \\
    0 & e^{i\frac{\pi}{4}}
    \end{pmatrix}\:.
\end{equation}
Note that since the $S$ gate $\bar{S}=\bar{T}^2$, the qRM code trivially supports a transversal logical $\bar{S}$ gate in terms of single-qubit $S/S^{\dagger}$ gates, respectively,
\begin{equation}
\bar{S}=S_1S^{\dagger}_2S_3S^{\dagger}_4S_5S^{\dagger}_6S_7S^{\dagger}_8S_9S^{\dagger}_{10}S_{11}S^{\dagger}_{12}S_{13}S^{\dagger}_{14}S_{15}\:.
\end{equation}
Even though the $[[15,1,3]]$ qRM supports a transversal non-Clifford gate, it does not support a transversal Haddamard gate $\bar{H}$, as this would violate Eastin-Knill theorem \cite{eastin2009restrictions}. Notably, both color codes described in this section achieve optimality between dimensionality and the Clifford hierarchy of the native transversal gates, saturating the bound proved for topological codes by the Bravyi-K{\"o}nig theorem \cite{bravyi2013}.

\section{Protocol}
\label{sec:protocol}

\begin{figure}[!ht]
\centering
\[
\begin{array}{c}
\centering
\quad \quad\quad\quad\quad\quad\Qcircuit @C=0.001em @R=.1em  @!{\lstick{\text{qRM:}\:\:\:\:|\bar{+}\rangle} &\gate{\bar{T}}   &\ctrl{1}   & \measuretab{M_{\bar{X}}} & \control \cw &  \\
\lstick{\text{Steane:}\:\:\:\:|\bar{0}\rangle} &\qw  &\targ & \qw & \gate{\bar{Z}} \cwx & \qw &\lstick{|\bar{T}\rangle}
}
\end{array}
\]
\caption{Protocol at the logical level. A qRM (quantum Reed-Muller) and a Steane codeblock are prepared in the $|\bar{+}\rangle$ and the $|\bar{0}\rangle$ state, respectively. The transversal logical $\bar{T}$ gate native to the qRM code is applied, creating a logical magic state $|\bar{T}\rangle=\bar{T}|\bar{+}\rangle$. The magic state is then teleported to the Steane code block by applying a logical $\overline{\text{CNOT}}$ gate, measuring the logical $\bar{X}$ operator and applying a logical $\bar{Z}$ correction in the case where $\bar{X}=-1$. At the end of the protocol, the logical magic state $|\bar{T}\rangle$ is prepared in the Steane code block.} \label{fig:logical_circuit} 
\end{figure}
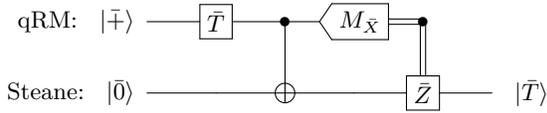

Our protocol aims to prepare a magic state encoded in the Steane code. At the logical level, our approach uses the standard one-bit teleportation protocol~\cite{Zhou2000} to teleport the magic state $|\bar{T}\rangle$ encoded in the qRM code to the Steane code, see Fig.~\ref{fig:logical_circuit}. 

We now describe each of the components in Fig.~\ref{fig:logical_circuit}. By strategically combining various existing fault-tolerant constructions from the literature~\cite{Bombin2006,Bombin2007,Chao2018,Chao2018a}, we succeed in giving an overall improvement compared to other known methods. Our main contribution is a novel protocol for the preparation of the $|\bar{+}\rangle$ state of the qRM code that uses parallel flagged syndrome extraction circuits with only a few extra ancillas and the utilization of a recently discovered $\overline{\text{CNOT}}$ gate.

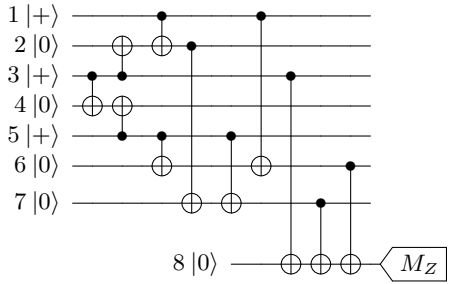
\begin{figure}[ht]
\[
\begin{array}{c}
\quad \quad \quad \quad \Qcircuit @C=.4em @R=.7em  {
\lstick{1\:|+\rangle}    &\qw      &\qw      &\qw &\ctrl{+1}&\qw      &\qw      &\qw      &\ctrl{+5}&\qw      &\qw      &\qw      &\qw  \\
\lstick{2\:|0\rangle}    &\qw      &\targ    &\qw &\targ    &\ctrl{+5}&\qw      &\qw      &\qw      &\qw      &\qw      &\qw      &\qw \\
\lstick{3\:|+\rangle}    &\ctrl{+1}&\ctrl{-1}&\qw &\qw      &\qw      &\qw      &\qw      &\qw      &\ctrl{+6}&\qw      &\qw      &\qw   \\
\lstick{4\:|0\rangle}    &\targ    &\targ    &\qw &\qw      &\qw      &\qw      &\qw      &\qw      &\qw      &\qw      &\qw      &\qw  \\
\lstick{5\:|+\rangle}    &\qw      &\ctrl{-1}&\qw &\ctrl{+1}&\qw      &\qw      &\ctrl{+2}&\qw      &\qw      &\qw      &\qw      &\qw \\
\lstick{6\:|0\rangle}    &\qw      &\qw      &\qw &\targ    &\qw      &\qw      &\qw      &\targ    &\qw      &\qw      &\ctrl{+3}&\qw  \\
\lstick{7\:|0\rangle}    &\qw      &\qw      &\qw &\qw      &\targ    &\qw      &\targ    &\qw      &\qw      &\ctrl{+2}&\qw      &\qw  \\
&\\
& & & & & & & \lstick{8\:|0\rangle}&\qw&\targ&\targ&\targ&\qw&\measuretab{M_Z}\\
}
\end{array}
\]
\caption{Preparation of the logical zero state $|\bar{0}\rangle$ in the Steane code. A non-fault-tolerant initialization involves preparing the physical qubits either in the zero $|0\rangle$ or plus $|+\rangle$ state, followed by a series of CNOT gates. A final verification step measures the logical $\bar{Z}$ operator and uses an ancilla qubit acting as a flag, detecting all weight-$2$ $X$ errors originating from a single $X$ error.} \label{fig:init_zero_Steane} 
\end{figure}

For completeness, we first describe the standard methods we employ. For the logical $\bar{T}$ gate of the qRM code, we apply the standard transversal gate (\ref{eq:Tgate})~\cite{Bravyi2005,Bombin2007}. For the logical-$X$ measurement necessary for the teleportation, we first measure all the qubits of the qRM code in the $X$ basis. Then, we compute the $X$ syndrome (see Table~\ref{tab:X_stab_qRM}) in postprocessing. At this point, we have a choice to either apply error correction, or instead postselect on measuring a trivial syndrome. We opt for the latter, since that yields a lower logical infidelity. After post-selecting for a trivial $X$ syndrome, we extract the value of the logical-$X$ operator $\bar{X}$, and we apply a logical-$Z$ correction on the Steane code in the case where $\bar{X}=-1$. The $|\bar{0}\rangle$ state preparation of the Steane code follows the known protocol of Refs.~\cite{Paetznick2012,goto2016minimizing,postler2022demonstration}; see Fig.~\ref{fig:init_zero_Steane}. The logical-$Z$ correction on the Steane code uses the standard weight-$3$ logical operator~(\ref{eq:logicalZSteane}). 

Lastly, since the final magic state is encoded in the Steane code, an additional round of flag-based error correction could be used to detect further errors as in Refs.~\cite{Chao2018,postler2022demonstration}, at the expense of increasing the overhead cost of the protocol. Given the depth of syndrome extraction circuits of Sec.~\ref{subsec:prep_plus}, we decided not to incorporate this variation.

\subsection{Preparation of $|\bar{+}\rangle$ for qRM code}
\label{subsec:prep_plus}
The most involved part of our protocol is the preparation of the $|\bar{+}\rangle$ for the qRM code. At a high level, our approach prepares the logical $|\bar{+}\rangle$ state using the encoding method of Ref.~\cite{Butt2024}, followed by an adapted version of flag-based error detection~\cite{Yoder2017,Chao2018,Chao2018a,Reichardt2020}. Our adapted flag-based error detection scheme employs a scheme inspired by Ref.~\cite{Chao2018}, which is appropriately modified to the qRM code. 

\begin{figure}[ht]
\[
\begin{array}{c}
\quad \quad \quad \Qcircuit @C=.001em @R=.45em  {
\lstick{1\:|0\rangle}    &\qw      &\qw      &\qw      &\qw      &\qw      &\qw      &\targ    &\qw      &\qw      &\qw      &\qw      &\qw      &\targ    &\qw      &\qw      &\qw      &\qw  &\qw  &\qw  &\qw  &\qw  &\qw  &\qw  &\qw  & \\
\lstick{2\:|0\rangle}    &\qw      &\targ    &\qw      &\qw      &\qw      &\qw      &\qw      &\qw      &\qw      &\targ    &\qw      &\qw      &\ctrl{-1}&\qw      &\qw      &\qw      &\targ&\qw  &\qw  &\qw  &\qw  &\qw  &\qw  &\qw  &     \\
\lstick{3\:|0\rangle}    &\qw      &\qw      &\qw      &\qw      &\qw      &\targ    &\qw      &\qw      &\qw      &\qw      &\qw      &\qw      &\qw      &\qw      &\targ    &\qw      &\qw  &\qw  &\qw  &\qw  &\qw  &\qw  &\qw  &\qw  &    \\
\lstick{4\:|0\rangle}    &\targ    &\qw      &\qw      &\qw      &\qw      &\ctrl{-1}&\qw      &\qw      &\qw      &\qw      &\targ    &\qw      &\qw      &\qw      &\qw      &\qw      &\qw  &\targ&\qw  &\qw  &\qw  &\qw  &\qw  &\qw  &  \\
\lstick{5\:|0\rangle}    &\qw      &\qw      &\qw      &\qw      &\qw      &\qw      &\qw      &\qw      &\qw      &\qw      &\qw      &\qw      &\targ    &\targ    &\qw      &\qw      &\qw  &\qw  &\qw  &\qw  &\qw  &\qw  &\qw  &\qw  & \\
\lstick{6\:|0\rangle}    &\qw      &\qw      &\qw      &\qw      &\qw      &\qw      &\qw      &\qw      &\targ    &\qw      &\qw      &\targ    &\qw      &\ctrl{-1}&\qw      &\targ    &\qw  &\qw  &\targ&\qw  &\qw  &\qw  &\qw  &\qw  & \\
\lstick{7\:|+\rangle}    &\ctrl{-3}&\ctrl{-5}&\targ    &\qw      &\qw      &\qw      &\qw      &\qw      &\qw      &\qw      &\qw      &\ctrl{-1}&\qw      &\qw      &\qw      &\qw      &\qw  &\qw  &\qw  &\qw  &\qw  &\qw  &\qw  &\qw  &  \\
\lstick{8\:|0\rangle}    &\qw      &\qw      &\qw      &\qw      &\targ    &\targ    &\ctrl{-7}&\targ    &\qw      &\qw      &\qw      &\qw      &\qw      &\qw      &\qw      &\qw      &\qw  &\qw  &\qw  &\targ&\qw  &\qw  &\qw  &\qw  &   \\
\lstick{9\:|+\rangle}    &\qw      &\qw      &\qw      &\qw      &\qw      &\qw      &\qw      &\ctrl{-1}&\qw      &\ctrl{-7}&\qw      &\ctrl{+1}&\qw      &\qw      &\qw      &\qw      &\qw  &\qw  &\qw  &\qw  &\qw  &\qw  &\qw  &\qw  &  \\
\lstick{10\:|0\rangle} &\qw      &\targ    &\qw      &\qw      &\qw      &\qw      &\qw      &\qw      &\qw      &\qw      &\qw      &\targ    &\qw      &\ctrl{+5}&\ctrl{-7}&\qw      &\qw  &\qw  &\qw  &\qw  &\targ&\qw  &\qw  &\qw  &  \\
\lstick{11\:|+\rangle} &\ctrl{+1}&\ctrl{-1}&\qw      &\targ    &\qw      &\qw      &\qw      &\ctrl{+4}&\qw      &\qw      &\ctrl{-7}&\qw      &\qw      &\qw      &\qw      &\qw      &\qw  &\qw  &\qw  &\qw  &\qw  &\qw  &\qw  &\qw  & \\
\lstick{12\:|0\rangle} &\targ    &\targ    &\qw      &\qw      &\qw      &\qw      &\ctrl{+3}&\qw      &\qw      &\qw      &\qw      &\qw      &\ctrl{-7}&\qw      &\qw      &\qw      &\qw  &\qw  &\qw  &\qw  &\qw  &\targ&\qw  &\qw  &  \\
\lstick{13\:|+\rangle} &\qw      &\ctrl{-1}&\qw      &\qw      &\ctrl{-5}&\qw      &\qw      &\qw      &\qw      &\qw      &\qw      &\qw      &\qw      &\qw      &\qw      &\ctrl{-7}&\qw  &\qw  &\qw  &\qw  &\qw  &\qw  &\qw  &\qw  &  \\  
\lstick{14\:|+\rangle} &\qw      &\qw      &\ctrl{-7}&\ctrl{-3}&\qw      &\ctrl{-6}&\qw      &\qw      &\ctrl{-8}&\qw      &\qw      &\qw      &\qw      &\qw      &\qw      &\qw      &\qw  &\qw  &\qw  &\qw  &\qw  &\qw  &\targ&\qw  & \\
\lstick{15\:|0\rangle} &\qw      &\qw      &\qw      &\qw      &\qw      &\qw      &\targ    &\targ    &\qw      &\qw      &\qw      &\qw      &\qw      &\targ    &\qw      &\qw      &\qw  &\qw  &\qw  &\qw  &\qw  &\qw  &\qw  &\qw  & \\
&\\
& & & & & & & & & & & & & & & & \lstick{16\:|+\rangle} &\ctrl{-15}&\ctrl{-13}&\ctrl{-11}&\ctrl{-9}&\ctrl{-7}&\ctrl{-5}&\ctrl{-3}&\measuretab{M_X}\\
}
\end{array}
\]
\caption{Preparation of the logical plus state $|\bar{+}\rangle$ in the qRM code. A non-fault-tolerant initialization involves preparing the physical qubits either in the zero $|0\rangle$ or plus $|+\rangle$ state, followed by a series of CNOT gates.} \label{fig:init_plus_state} 
\end{figure}
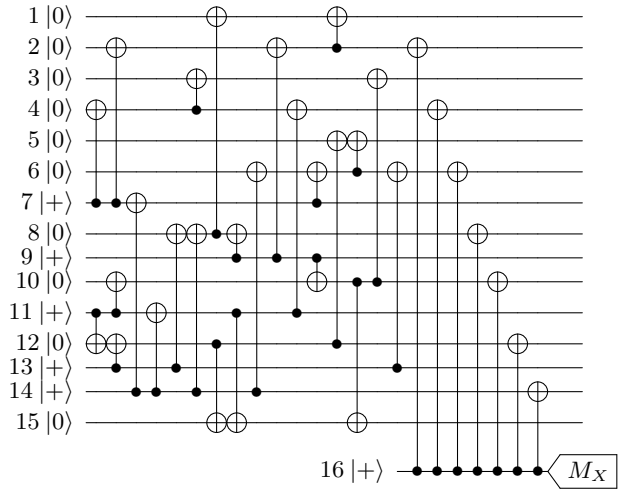

The first step of this protocol is described in Fig.~\ref{fig:init_plus_state}. The final verification step measures the logical $\bar{X}$ operator and uses an ancilla qubit acting as a flag, detecting all uncorrectable weight-$2$ $Z$ errors originating from a single $Z$ error.

At this point, the state prepared may have additional $X$ errors. If we do not correct them, they will propagate through the transversal $T$ gate to a $XS$ error, which will lead to an error in the final measurement $M_{\bar{X}}$ of the qRM code. In order to mitigate this effect, one must measure the $Z$-type syndromes. Such measurements should be chosen carefully. If we were to measure each $Z$-type syndrome using the standard approach (i.e., using a single ancilla and four CNOTs), a single fault can propagate to weight-2 errors in the data block.

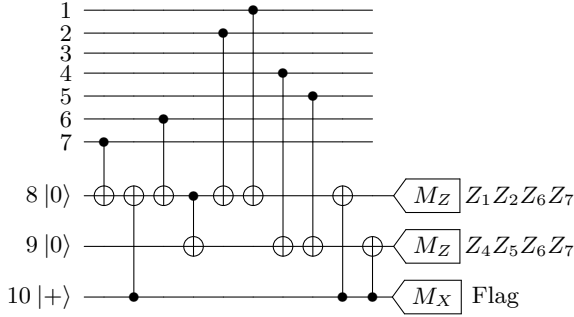
\begin{figure}[ht]
\[
\begin{array}{c}
\Qcircuit @C=.4em @R=.7em  {
\lstick{1}              &\qw      &\qw      &\qw      &\qw      &\qw      &\ctrl{+8}&\qw      &\qw      &\qw      &\qw      &\\
\lstick{2}              &\qw      &\qw      &\qw      &\qw      &\ctrl{+7}&\qw      &\qw      &\qw      &\qw      &\qw      &\\
\lstick{3}              &\qw      &\qw      &\qw      &\qw      &\qw      &\qw      &\qw      &\qw      &\qw      &\qw      &\\
\lstick{4}              &\qw      &\qw      &\qw      &\qw      &\qw      &\qw      &\ctrl{+6}&\qw      &\qw      &\qw      &\\
\lstick{5}              &\qw      &\qw      &\qw      &\qw      &\qw      &\qw      &\qw      &\ctrl{+5}&\qw      &\qw      &\\
\lstick{6}              &\qw      &\qw      &\ctrl{+3}&\qw      &\qw      &\qw      &\qw      &\qw      &\qw      &\qw      &\\
\lstick{7}              &\ctrl{+2}&\qw      &\qw      &\qw      &\qw      &\qw      &\qw      &\qw      &\qw      &\qw      &\\
&\\
\lstick{8\:|0\rangle}   &\targ    &\targ    &\targ    &\ctrl{+1}&\targ    &\targ    &\qw      &\qw      &\targ    &\qw      &\measuretab{M_Z}&\quad \quad \quad \quad Z_1Z_2Z_6Z_7 \\
\lstick{9\:|0\rangle}   &\qw      &\qw      &\qw      &\targ    &\qw      &\qw      &\targ    &\targ    &\qw      &\targ    &\measuretab{M_Z}&\quad \quad \quad \quad Z_4Z_5Z_6Z_7 \\
\lstick{10\:|+\rangle}&\qw      &\ctrl{-2}&\qw      &\qw      &\qw      &\qw      &\qw      &\qw      &\ctrl{-2}&\ctrl{-1}&\measuretab{M_X}&\quad \quad  \text{Flag}\\
}
\end{array}
\]
\caption{Simultaneous flagged syndrome extraction circuit for two $Z$-type stabilizers in the Steane code. An additional qubit prepared in the plus state $|+\rangle$ acts as a flag and detects higher weight data $Z$ errors originating from arbitrary single-qubit ancilla $Z$ errors.} \label{fig:2_Z_stab_Steane} 
\end{figure}

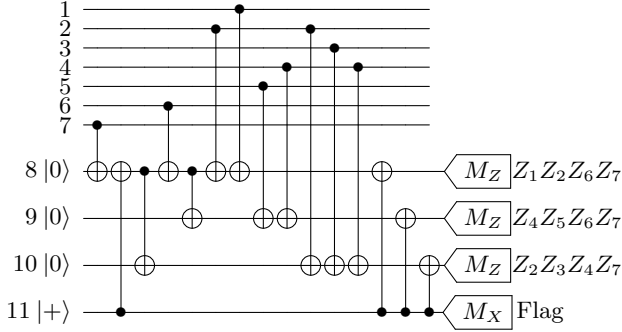
\begin{figure}[ht]
\[
\begin{array}{c}
\Qcircuit @C=.15em @R=.55em  {
\lstick{1}              &\qw      &\qw      &\qw      &\qw      &\qw      &\qw      &\ctrl{+8}&\qw      &\qw      &\qw      &\qw      &\qw      &\qw      &\qw      &\qw      &\\
\lstick{2}              &\qw      &\qw      &\qw      &\qw      &\qw      &\ctrl{+7}&\qw      &\qw      &\qw      &\ctrl{+9}&\qw      &\qw      &\qw      &\qw      &\qw      &\\
\lstick{3}              &\qw      &\qw      &\qw      &\qw      &\qw      &\qw      &\qw      &\qw      &\qw      &\qw      &\ctrl{+8}&\qw      &\qw      &\qw      &\qw      &  \\
\lstick{4}              &\qw      &\qw      &\qw      &\qw      &\qw      &\qw      &\qw      &\qw      &\ctrl{+6}&\qw      &\qw      &\ctrl{+7}&\qw      &\qw      &\qw      & \\
\lstick{5}              &\qw      &\qw      &\qw      &\qw      &\qw      &\qw      &\qw      &\ctrl{+5}&\qw      &\qw      &\qw      &\qw      &\qw      &\qw      &\qw      &\\
\lstick{6}              &\qw      &\qw      &\qw      &\ctrl{+3}&\qw      &\qw      &\qw      &\qw      &\qw      &\qw      &\qw      &\qw      &\qw      &\qw      &\qw      &\\
\lstick{7}              &\ctrl{+2}&\qw      &\qw      &\qw      &\qw      &\qw      &\qw      &\qw      &\qw      &\qw      &\qw      &\qw      &\qw      &\qw      &\qw      &\\
&\\
\lstick{8\:|0\rangle}   &\targ    &\targ    &\ctrl{+2}&\targ    &\ctrl{+1}&\targ    &\targ    &\qw      &\qw      &\qw      &\qw      &\qw      &\targ    &\qw      &\qw      &\measuretab{M_Z}&\quad \quad \quad \quad \: Z_1Z_2Z_6Z_7\\
\lstick{9\:|0\rangle}   &\qw      &\qw      &\qw      &\qw      &\targ    &\qw      &\qw      &\targ    &\targ    &\qw      &\qw      &\qw      &\qw      &\targ    &\qw      &\measuretab{M_Z}&\quad \quad \quad \quad \:Z_4Z_5Z_6Z_7\\
\lstick{10\:|0\rangle}&\qw      &\qw      &\targ    &\qw      &\qw      &\qw      &\qw      &\qw      &\qw      &\targ    &\targ    &\targ    &\qw      &\qw      &\targ    &\measuretab{M_Z}&\quad \quad \quad \quad \:Z_2Z_3Z_4Z_7 \\
\lstick{11\:|+\rangle}&\qw      &\ctrl{-3}&\qw      &\qw      &\qw      &\qw      &\qw      &\qw      &\qw      &\qw      &\qw      &\qw      &\ctrl{-3}&\ctrl{-2}&\ctrl{-1}&\measuretab{M_X}&\quad \quad  \text{Flag}\\
}
\end{array}
\]
\caption{Simultaneous flagged syndrome extraction circuit for three $Z$-type stabilizers in the Steane code. An additional qubit prepared in the plus state $|+\rangle$ acts as a flag and detects higher weight data $Z$-errors originating from arbitrary single qubit ancilla $Z$-errors.} \label{fig:3_Z_stab_Steane} 
\end{figure}

We can avoid this problem by using the parallel flag-based syndrome-extraction circuits for the \textit{Steane code}~\cite{Chao2018}, adapted to the qRM code. In that paper, the authors introduced two types of syndrome extraction circuits, one extracting two syndromes at once and the other extracting three syndromes at once; see Figs.~\ref{fig:2_Z_stab_Steane} and~\ref{fig:3_Z_stab_Steane}. In the qRM code, there are ten independent $Z$-type stabilizers, see Table~\ref{tab:Z_stab_qRM}, which we can organize as follows. First, we can measure $(Z_8Z_{12}Z_{13}Z_{15}$, $Z_5Z_6Z_{12}Z_{13})$ and $(Z_3Z_4Z_{10}Z_{11}$, $Z_4Z_5Z_{11}Z_{15})$ using  Fig.~\ref{fig:2_Z_stab_Steane}. Then we can measure $(Z_1Z_2Z_6Z_7$, $Z_2Z_3Z_4Z_7$, $Z_4Z_5Z_6Z_7)$ and $(Z_8Z_9Z_{13}Z_{14}$, $Z_9Z_{10}Z_{11}Z_{14}$, $Z_{11}Z_{12}Z_{13}Z_{14})$ using Fig.~\ref{fig:3_Z_stab_Steane}. The advantage of using this new scheme is that instead of employing ten ancilla qubits to flag each syndrome extraction circuit individually, we reduce the number to four. Moreover, the number of time steps needed to run a full syndrome extraction cycle is also reduced, which is desirable for experimental demonstrations. Flags and syndromes are postselected to be $+1$. We remark that there are other sets of stabilizers one may choose to measure. \footnote{It turns out that four is the minimum number of $Z$-type stabilizers needed to \textit{detect} all propagated weight-$2$ or higher $X$ errors after the initialization of Fig.~\ref{fig:init_plus_state} For instance, a set of plaquette stabilizers that satisfy this condition is $\{p_2,p_3,p_8,p_{13}\}$. In an experiment, this reduction can decrease the space-time cost of the protocol (see Sec.~\ref{subsec:spacetime_cost}). We thank an anonymous referee for suggesting to inspect this property.}. We leave such studies for future work.

\subsection{Transversal CNOT gate}
\label{subsec:transversal_gate}

The logical $\overline{\text{CNOT}}$ gate in Fig.~\ref{fig:logical_circuit} can be implemented using a transversal gate. With respect to the convention described in Figs.~\ref{fig:Steane} and~\ref{fig:qRM}, we can apply the following operation:
\begin{equation}
    \overline{\text{CNOT}} = \bigotimes_{i=1}^7 \text{CNOT}_i\:, \label{eq:logical_cnot}
\end{equation}
where $\text{CNOT}_i$ is the CNOT gate whose control and target is the $i$'th qubit of the qRM and Steane code, respectively. The qubits belonging to the qRM code that participate in the $\text{CNOT}$ gate span the bottom face of the tetrahedron of Fig.~\ref{fig:qRM}. In addition, the other orientation of the CNOT
gate not used in this protocol is not available via the transversal
method of (\ref{eq:logical_cnot}).

This gate preserves the stabilizer group and satisfies the following relations:
\begin{equation}
\begin{aligned}
    \overline{\text{CNOT}} \bar{X}_{\text{qRM}}\overline{\text{CNOT}} &=  \bar{X}_{\text{qRM}} \bar{X}_{\text{S}} \\
    \overline{\text{CNOT}} \bar{Z}_{\text{S}}\overline{\text{CNOT}} &=  \bar{Z}_{\text{qRM}} \bar{Z}_{\text{S}}
\end{aligned}
\end{equation}
up to stabilizers, where $\bar{X}_{\text{qRM}}$ and $\bar{Z}_{\text{qRM}}$ are logical $X$ and $Z$ operators of the qRM code, (\ref{eq:logicalXqRM}) and (\ref{eq:logicalZqRM}); similarly,  $\bar{X}_{\text{S}}$ and $\bar{Z}_{\text{S}}$ are logical $X$ and $Z$ operators of the Steane code, (\ref{eq:logicalXSteane}) and (\ref{eq:logicalZSteane}). Thus, Eq.~\eqref{eq:logical_cnot} is a logical CNOT from the qRM code to the Steane code. A generalization of this construction to the entire 3D and 2D color code family is discussed in Ref.~\cite{Heuben2024}.

\subsection{Space-time Cost}
\label{subsec:spacetime_cost}

In this section, we summarize the overall space-time cost of our protocol. The majority of the cost comes from $|\bar{+}\rangle$ state preparation (Fig.~\ref{fig:init_plus_state}) and the syndrome extraction circuits (Figs.~\ref{fig:2_Z_stab_Steane} and~\ref{fig:3_Z_stab_Steane}). Overall, we estimate the total space-time volume to be at most $\approx 1.3\times 10^3$; see Table~\ref{tab:spacetime_cost} for a breakdown. 

It is interesting to note that the overall space-time cost is significantly smaller than methods based on magic state distillation. For instance, consider a recent distillation protocol based on 2D color codes, studied in Ref.~\cite{Lee2024}. In Fig.~14 of Ref.~\cite{Lee2024}, the authors report an overall space-time cost needed to achieve a target infidelity. At a physical error rate of $10^{-3}$, for an infidelity of order $10^{-6}$, there is no distillation-based protocol with a space-time cost less than $10^6$. On the other hand, our protocol achieves an infidelity of $5.1\times 10^{-7}$ (with a postselection approach) using a space-time volume of $\approx 1.3\times 10^3$, which is a significant improvement.

Furthermore, we note that our estimate of $1.3\times 10^3$ is likely an overestimate. For instance, the circuits of Fig.~\ref{fig:3_Z_stab_Steane} could be implemented in parallel in our protocol. So, the space-time cost of the most resource-intensive individual part of the protocol in principle can be at least halved. 

\begin{table*}[ht]
    \centering
    \begin{tabular}{|c|c|c|c|c|c|}
    \hline
        Protocol&  Volume  & $2$-qubit gates & Measurements & Depth\\
        \hline
       $|\bar{0}\rangle$ State preparation (Fig.~\ref{fig:init_zero_Steane}) & $75$ & $11$ & $1$ & $10$\\
       $|\bar{+}\rangle$ State preparation (Fig.~\ref{fig:init_plus_state}) & $309$ & $32$ & $1$& $20$ \\
       Syndrome extraction (Fig.~\ref{fig:2_Z_stab_Steane}) & $360$ & $20$ & $6$ & $20$\\
       Syndrome extraction (Fig.~\ref{fig:3_Z_stab_Steane}) & $532$ & $30$ & $8$ & $28$\\
       Transversal $T$ gate & $15$ & $-$ & $-$ & $1$\\
       Transversal CNOT gate & $22$ & $7$ & $-$ & $1$\\
       Logical-$X$ measurement & $22$ & $-$ & $15$ & $1$ \\
       \hline
       Total & $1335$ & $100$ & $31$ & $81$\\
       \hline
    \end{tabular}
    \caption{Summary of the total space-time cost estimate of the protocol, together with the number of $2$-qubit CNOT gates, single-qubit destructive measurements, and circuit depth. The space-time cost (volume) is defined as the number of qubits times the number of time steps. For the syndrome extraction circuits, the estimates contemplate two uses for each type of circuit (see Sec.~\ref{subsec:prep_plus}).}
    \label{tab:spacetime_cost}
\end{table*}


\section{Simulation}
\label{sec:simulation}

The protocol described in Sec.~\ref{sec:protocol} uses $22$ data qubits and $16$ ancilla qubits. If the ancilla qubits are reused, it is possible to reduce the ancilla qubit requirements to four, which is bottlenecked by the flag circuit in Fig.~\ref{fig:3_Z_stab_Steane}. Even so, an exact \emph{mixed state} simulation of $26$ qubits is challenging.

As such, we resort to a Monte Carlo sampling-based approach. We keep track of the Pauli errors, which after passing through $\bar{T}$, become Clifford errors. Thus, by studying an \emph{ensemble} of circuits, each suffering from a random Clifford error, we can simulate the performance of our protocol. This is a simulation technique akin to an extended stabilizer simulator, which effectively incorporates the non-Clifford $T$ gate.

To incorporate the stochastic Clifford errors into the simulation, we first initialize the noisy logical plus state $|\bar{+}\rangle$ in the qRM code as in Fig.~\ref{fig:logical_circuit} and extract the Pauli error frame $P$, 
\begin{equation}
|\bar{+}\rangle=P|\bar{+}\rangle_{\text{noiseless}}\:,
\label{eq:pauli_frame}
\end{equation}
where $|\bar{+}\rangle_{\text{noiseless}}$ is the noiseless plus state. Since the $T$ gate has to be applied to the noisy plus state,
\begin{equation}
    \bar{T}|\bar{+}\rangle=(\bar{T}P\bar{T}^{\dagger})\bar{T}|\bar{+}\rangle_{\text{noiseless}}=C(P)|\bar{T}\rangle_{\text{noiseless}}\:.
\label{eq:magic_qrm_noisy}
\end{equation}
The state $|\bar{T}\rangle_{\text{noiseless}}=\bar{T}|\bar{+}\rangle_{\text{noiseless}}$ is a noiseless magic state in the qRM code and $C(P)=\bar{T}P\bar{T}^{\dagger}$ is the Clifford error frame. If $P \sim \bigotimes_{i=1}^{15}X^{a_i}_iZ^{b_i}_i$ for $a_i,b_i \in \{0,1\}$, then the Clifford error frame becomes
\begin{equation}
    C(P)\sim \left(\bigotimes_{i\:\text{odd}}S^{a_i}_i\right)\left(\bigotimes_{i\:\text{even}}S^{\dagger a_i}_i\right) P\:,
\label{eq:CP}
\end{equation}
up to a global phase. The ordering of the gates appearing in (\ref{eq:CP}) is important, with the Pauli error frame $P$ (\ref{eq:pauli_frame}) acting first, followed by the phase gates. We discuss the significance of the appropriate ordering in the last paragraph of Appendix~\ref{app:sim_details}.

After updating the Pauli errors to Clifford errors, we effectively have an ensemble of encoded $|\bar{T}\rangle_{\text{noiseless}}$ states of the qRM code followed by a stochastic Clifford noise. Note that for a magic state $|\bar{T}\rangle$, the density matrix $|\bar{T}{\rangle}{\langle} \bar{T}|$ can be written as a linear combination of pure stabilizer states:
\begin{equation}
\begin{aligned}
    |\bar{T}{\rangle}{\langle} \bar{T}| &=\frac{1}{2} I+\frac{\sqrt{2}}{4} (|\bar{+}{\rangle}{\langle} \bar{+}|-|\bar{-}{\rangle}{\langle} \bar{-}|)\\
    &+ \frac{\sqrt{2}}{4}(|\bar{Y}_+{\rangle}{\langle} \bar{Y}_+| - |\bar{Y}_-{\rangle}{\langle} \bar{Y}_-|).
\end{aligned}
\label{eq:non-convex}
\end{equation}
All these states can be implemented by applying logical transversal gates to the logical plus state in the qRM code: $|\bar{-}\rangle=\bar{Z}|\bar{+}\rangle$, $|\bar{Y}_+\rangle=\bar{S}|\bar{+}\rangle$ and $|\bar{Y}_-\rangle=\bar{S}^{\dagger}|\bar{+}\rangle$. For each of the stabilizer states appearing in Eq.~\eqref{eq:non-convex}, we can estimate the fidelity between the logical noiseless magic state in the Steane code $|\bar{T}\rangle_{\text{Steane}}$ (after teleportation)  and the given noisy final state $\bar{\rho}$ using the following relation:
\begin{equation}
    \langle \bar{T} | \bar{\rho} |\bar{T}\rangle_{\text{Steane}} = \frac{1}{2} + \frac{\sqrt{2}}{4} (\langle \bar{X}\rangle_{\bar{\rho}} + \langle \bar{Y}\rangle_{\bar{\rho}}),\label{eq:fidelity_formula}
\end{equation}
where $\langle\bar{X}\rangle_{\bar{\rho}}$ and $\langle \bar{Y}\rangle_{\bar{\rho}}$ are the expectation values of the logical $X$ and $Y$ operators evaluated with respect to the code state $\bar{\rho}$, i.e., $\langle \mathcal{O}\rangle_{\bar{\rho}}=\text{Tr}(\mathcal{O} \bar{\rho})$.

In our simulation, the state $\bar{\rho}$ is chosen to be the (normalized) postselected state that we obtain at the very end of the protocol. Because the postselection probability is identical for every stabilizer state appearing in Eq.~\eqref{eq:non-convex}, we can estimate Eq.~\eqref{eq:fidelity_formula} by taking the expectation value of the logical $X$ and $Y$ operator of the Steane code only for the post-selected events. Indeed, by combining (\ref{eq:non-convex}) and (\ref{eq:fidelity_formula}), the fidelity can be conveniently written as 
\begin{equation}
    \langle \bar{T} | \bar{\rho} |\bar{T}\rangle_{\text{Steane}} = \frac{1}{2} + \frac{1}{8}\Delta \:,
\label{eq:fidelity_practical1}
\end{equation}
where
\begin{equation}
\begin{split}
    \Delta & = \Big(\langle \bar{X}\rangle_{|\bar{+}\rangle}-\langle \bar{X}\rangle_{|\bar{-}\rangle}+\langle \bar{X}\rangle_{|\bar{Y}_+\rangle}-\langle \bar{X}\rangle_{|\bar{Y}_-\rangle}\Big)\\
    & \quad \quad +\Big(\langle \bar{Y}\rangle_{|\bar{+}\rangle}-\langle \bar{Y}\rangle_{|\bar{-}\rangle}+\langle \bar{Y}\rangle_{|\bar{Y}_+\rangle}-\langle \bar{Y}\rangle_{|\bar{Y}_-\rangle}\Big)\:.
\end{split}
\label{eq:fidelity_practical2}
\end{equation}
A proof of (\ref{eq:fidelity_practical1}) and (\ref{eq:fidelity_practical2}) can be found in Appendix~\ref{app:sim_details}. We stress that the described simulation method is well suited for our protocol because we only employ once a logical transversal $T$ gate, thus, it is applied in a single layer at the physical level. However, in a more general scenario where two or more non-Clifford gates act on some subset of qubits (as in Refs.~\cite{goto2016minimizing,Chamberland2020}), it is unclear if our proposed method is applicable. This is because Pauli errors become Clifford errors after the first non-Clifford gate, and subsequently non-Clifford errors after the remaining non-Clifford gates.

We quantify the quality of the magic state using two different methods. In the first approach, we compute the fidelity using Eqs.~\eqref{eq:fidelity_practical1} and \eqref{eq:fidelity_practical2} assuming that at the final step we perform logical measurement. That is, we measure all the qubits in the $X$-, or $Y$- basis and obtain the syndrome information from the measurement outcomes (recall that in the Steane code, since the stabilizer generators have weight-$4$, it is possible to get the $Z$ syndrome by multiplying the $X$ and $Y$ syndromes, respectively). From the syndrome information, a correction is deduced using a look-up table, after which we obtain the logical measurement outcome. This approach roughly quantifies the quality of the magic state encoded in the Steane code. 

In the second approach, in the last step we postselect on measuring a trivial syndrome. This can be used as a proxy for the lowest error rate one can achieve if one were to later enlarge the code to a larger one whilst post-selecting~\cite{Itogawa2024,Gidney2024}. However, as pointed out in Ref.~\cite{Gidney2024}, the logical error rate obtained after such enlargement process is often higher than the one obtained prior to that stage. How the logical error rate changes under such a process is beyond the scope of this paper, and we leave it for a future work.

\subsection{Noise model}
\label{subsec:noise_model}

We use the noise model in which every operation is followed or preceded by a noisy process; we use the uniform depolarizing noise model of strength $p$. For state preparation, we assume that there is a probability $p$ of flipping the initial state. Hence, we model the preparation of a $|0\rangle$ state as perfect $|0\rangle$ followed by a bit flip $X$ applied with probability $p$. Similarly, the preparation of a $|+\rangle$ state is modeled as a perfect $|+\rangle$ state followed by a phase flip $Z$ with probability $p$.

For measurements, we follow a similar recipe as for state preparation. We model a $Z$-type measurement as a bit flip $X$ applied with probability $p$, followed by a perfect $Z$-type measurement. Similarly, we model an $X$-type measurement as a phase flip $Z$ applied with probability $p$, followed by a perfect $X$-type measurement.

For single- and two-qubit gates, we model them as perfect gates followed by the depolarizing error model. For $1$-qubit gates, the depolarizing error model leaves the state unaltered with probability $1-p$, and applies a Pauli error $P\in\{X,Y,Z\}$ (denoted as $\mathcal{P}_1$) with probability $\frac{p}{3}$. For 2-qubit gates,  the depolarizing error model leaves the state unaltered with probability $1-p$, and applies a Pauli error $P\otimes Q$ with $P,Q\in\{I,X,Y,Z\}$ (denoted as $\mathcal{P}_2$) and $P\otimes Q \neq I \otimes I$ with probability $\frac{p}{15}$. Specifically, these channels can be written as follows 
\begin{equation}
\mathcal{D}_1(\rho)=(1-p)\rho+\sum_{P\in\ \mathcal{P}_1/\{I\}}\frac{p}{3}\:P\rho P\:,
\end{equation}
\begin{equation}
\quad \quad\mathcal{D}_2(\rho)=(1-p)\rho +\sum_{P\in\ \mathcal{P}_2/\{I\otimes I\}}\frac{p}{15}\:P\rho P\:.
\end{equation}
Here, $\mathcal{D}_1$ and $\mathcal{D}_2$ are a single- and two-qubit depolarizing noise model, respectively.

Regarding idling errors, we do not consider them in our simulations. Given the depths of the different circuits present in our protocol, their contribution might not be entirely insignificant in a experimental demonstration, which we ultimately expect to give a conclusive assessment of the protocol. We plan to study this effect in more detail in a future work. 

In order to compare our findings with other state-of-the-art results in the literature, some simulations are done with the multiparameter depolarizing error model too, see Table~\ref{tab:comparisson2}. This model is defined similarly as the uniform counterpart, with the only difference that each type of error can be tuned independently, and they are parametrized by $p_i$ initialization error, $p_m$ measurement error, and $p_1$ and $p_2$ single- and two-qubit error gate.

\subsection{Simulation Methods}
\label{subsec:sim_methods}

Our simulations are done using the stabilizer simulator package Stim~\cite{Gidney2021} for Python. Specifically, using Pauli frames obtained from the functions $\texttt{FlipSimulator()}$ and $\texttt{peek\_pauli\_flips()}$, we generated random instances of Clifford errors, each of which are later incorporated to the full circuit using again Stim. For a given simulation, we run $N_{\text{tot}}$ Monte Carlo shots, and postselected $N_{\text{post}}$ of them. Hence, the acceptance rate of the protocol is given by 
\begin{equation}
p_{\text{accept}}=\frac{N_{\text{post}}}{N_{\text{tot}}}\:.
\label{eq:post_accept}
\end{equation}
For later reference, the postselection rejection probabilities are defined as the complement of the acceptance rate (\ref{eq:post_accept}), in other words ,
\begin{equation}
    p_{\text{reject}}=1-p_{\text{accept}}\:.
\end{equation} 
Similarly, the infidelity of the magic state quantifies the logical failure and is defined as the complement of the infidelity (\ref{eq:fidelity_practical1}) and (\ref{eq:fidelity_practical2}),
\begin{equation}
  p_{\text{fail}} = 1-\langle \bar{T}| \bar{\rho}|\bar{T}\rangle_{\text{Steane}}  \:.
\end{equation}

The total number of shots $N_{\text{tot}}$ ranges from $10^7$ up to $3\times 10^8$. To compute the uncertainty of the fidelity, we define first the uncertainty of the expectation values $ \langle \bar{P}\rangle_{\bar{\rho}}$ for $P=X,Y$, which is given by
\begin{equation}
    \sigma_{P,\bar{\rho}}=\sqrt{\frac{\text{Var}(\langle \bar{P}\rangle_{\bar{\rho}})}{N_{\text{post}}}}\:.
\end{equation}
We observe that at low error rates, the major contribution to the variance comes from the intrinsic probabilistic nature of the measurement of an eigenstate of a Pauli $P$ in another basis $P'$. More precisely, this occurs when computing in (\ref{eq:fidelity_practical1}) and  (\ref{eq:fidelity_practical2}) the expected values  $\langle \bar{X}\rangle_{|\bar{Y}_{\pm}\rangle}$ and $\langle \bar{Y}\rangle_{|\bar{\pm}\rangle}$. In a noiseless scenario, we can model the measurement outcomes as a Bernoulli distribution that returns $+1$ and $-1$ with even probability $1/2$, which has variance $1$, thus, yielding an error of  $\sigma_{P,\bar{\rho}} \gtrsim 1/\sqrt{N_{\text{post}}}$ for the aforementioned four expected values. Finally, taking into account normalization factors in (\ref{eq:fidelity_practical1}) and  (\ref{eq:fidelity_practical2}), this implies that the propagated fidelity error is roughly given by $\sigma_{ \langle \bar{T} | \bar{\rho} |\bar{T}\rangle_{\text{Steane}}} \gtrsim 1/(4\sqrt{N_{\text{post}}})$.

\subsection{Results}
\label{subsec:results}

As we stated before in Sec.~\ref{sec:simulation}, we consider two approaches. The first approach performs error correction when measuring the logical $X$ and $Y$ used in Eqs.~(\ref{eq:fidelity_practical1}) and  (\ref{eq:fidelity_practical2}). The second approach uses postselection on measuring trivial syndrome at this step.

\begin{figure}[!ht]
    \centering
    \includegraphics[width=0.48\textwidth]{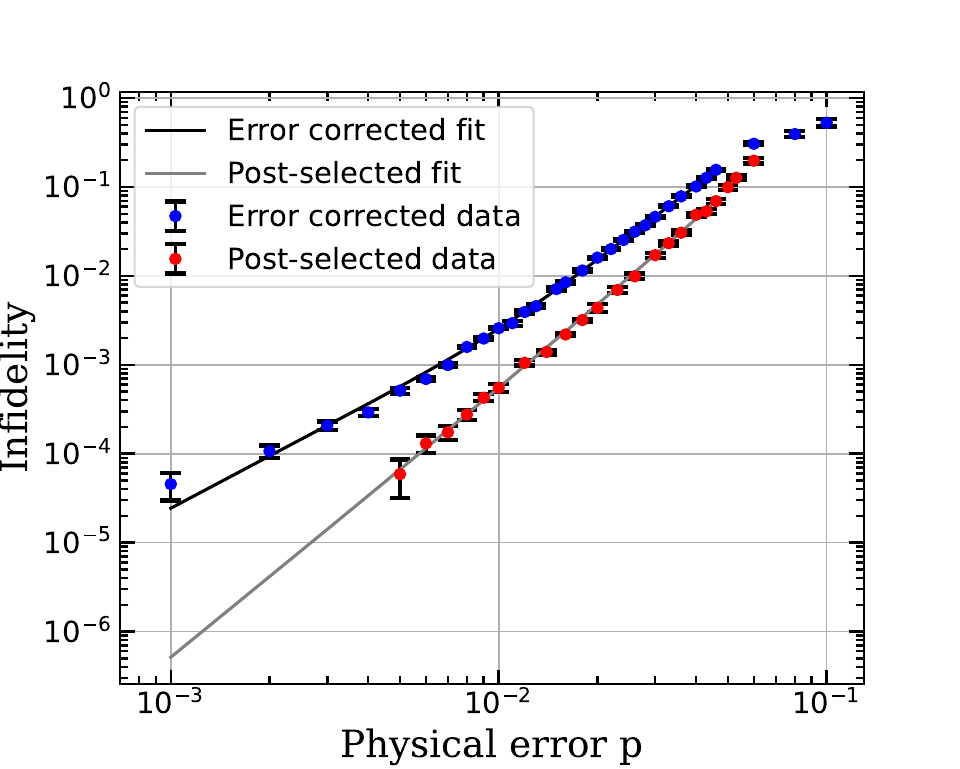}
    \caption{Infidelity of magic states $|\bar{T}\rangle$ as a function of physical error rate for a uniform depolarizing error model (excluding idling errors) of strength $p$. The polynomial fits with their respective parameters are explicitly given in Eq.~(\ref{eq:infidelity_fit}).}
    \label{fig:infidelity_plot}
\end{figure}

\begin{figure}[!ht]
    \centering
    \includegraphics[width=0.48\textwidth]{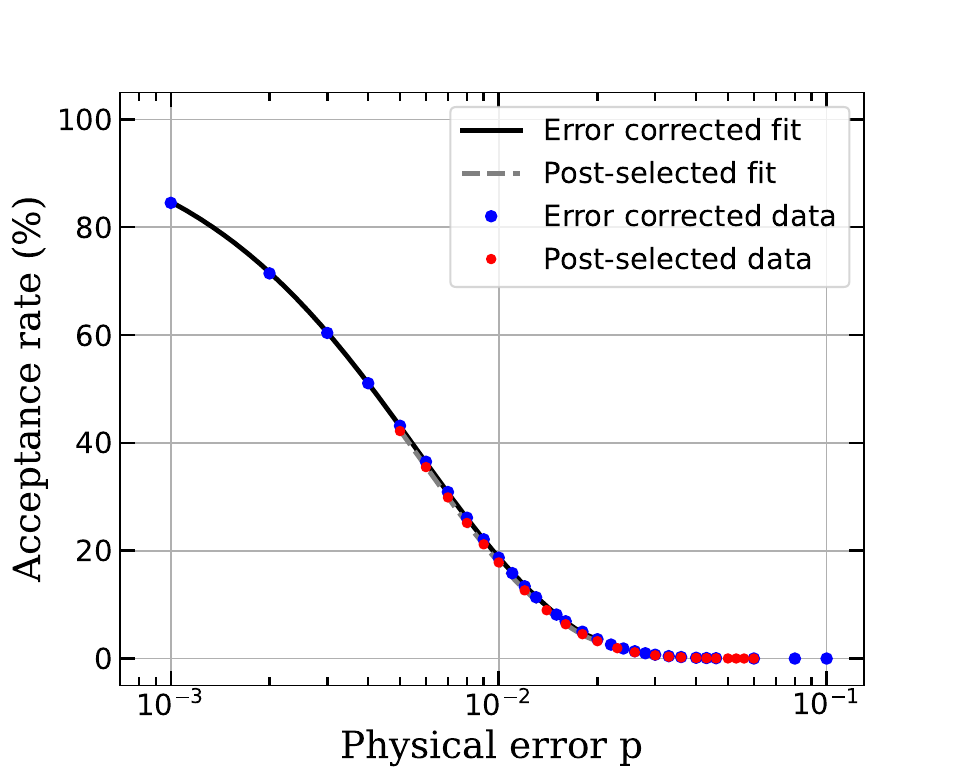}
    \caption{Acceptance rate of magic states $|\bar{T}\rangle$ as a function of physical error rate for a uniform depolarizing error model (excluding idling errors) of strength $p$. The polynomial fits with their respective parameters are explicitly given in Eq.~(\ref{eq:acceptance_fit}). A plot of the difference between acceptance rates in logarithmic scale can be found in Fig.~\ref{fig:difference_plot}.}
\label{fig:acceptance_plot}
\end{figure}

The results of the simulation for both approaches are plotted in Figs.~\ref{fig:infidelity_plot} and~\ref{fig:acceptance_plot}. In the first and the second approach, our circuit can detect all faults of weight $1$ and $2$, respectively; this difference comes from the final logical measurement stage. We have fitted the logical infidelity and the postselection probability to a polynomial. 
From the fitted polynomial for the infidelity we obtained at leading order, 
\begin{equation}
    p_{\text{fail}} \approx
    \begin{cases}
        25.5p^2 - 1266.7p^3+ 165884.4p^4 \\

        \quad \quad \quad-4339823.2p^5 +39742917.9 p^6\\
    \quad\quad\quad\quad\quad\quad\quad\quad\text{(error correction),}\\
        508.5p^3+4477.1p^4 \\
        \quad\quad\quad\quad\quad\quad\quad\quad \text{(postselection).}\\
    \end{cases}
\label{eq:infidelity_fit}
\end{equation}

In addition, at leading order the polynomial fits for the rejection rates yield,
\begin{equation}
    p_{\text{reject}} \approx
    \begin{cases}
        164.6p - 12362.8p^2+477206.9p^3  \\
        \quad \quad \quad \quad \quad\quad-7518868.5p^4        \\
    \quad\quad\quad\quad\quad\quad\quad\quad\text{(error correction),}\\
        166.4p - 12352.7p^2+462706.3p^3  \\
        \quad \quad \quad \quad \quad\quad-7014618.5p^4        \\\quad\quad\quad\quad\quad\quad\quad\quad \text{(postselection).}\\
    \end{cases}
\label{eq:acceptance_fit}
\end{equation}

\begin{figure}[!ht]
    \centering
    \includegraphics[width=0.48\textwidth]{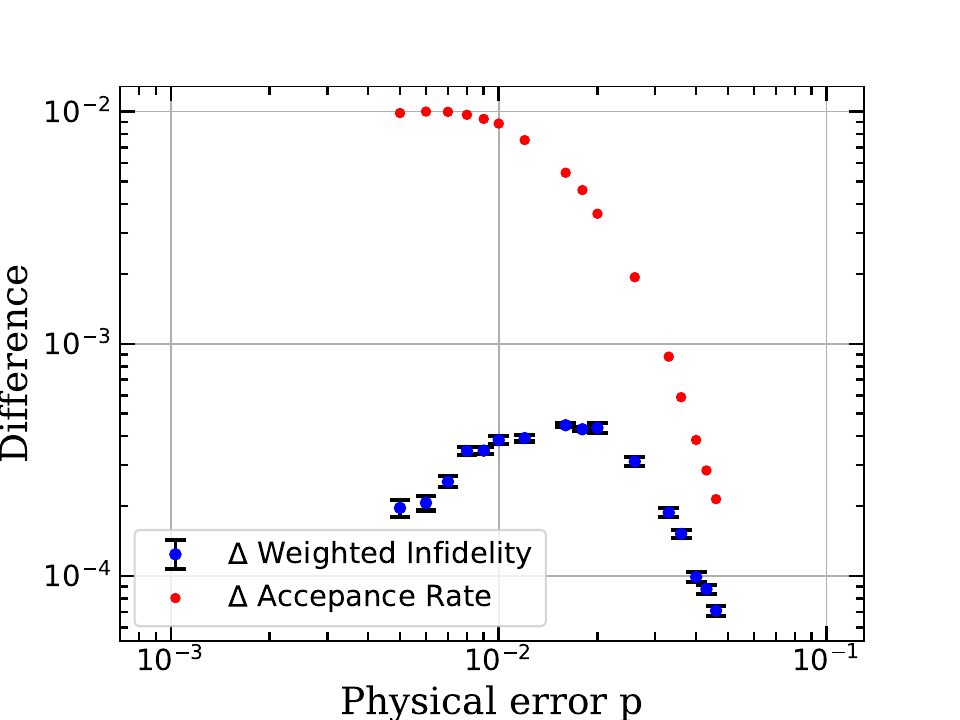}
    \caption{Plot of the difference between acceptance rates (\ref{eq:dif_acceptance}) (in red) and weighted infidelities (\ref{eq:weigth_inf}) (in blue) for the error-correction and postselection analysis, as a function of physical error rate for a uniform depolarizing error model (excluding idling errors) of strength $p$. Note that the vertical axis is in logarithmic scale and that the acceptance rate is not expressed in percent.}
\label{fig:difference_plot}
\end{figure}

The acceptance rate plots in Fig.~\ref{fig:acceptance_plot} suggest a small difference between the error-corrected and the postselected simulation results. Hence, in Fig.~\ref{fig:difference_plot} we show a plot in logarithmic scale of the difference between the acceptance rate using error correction $p_{\text{accept}}^{EC}$ and postselection/error-detection $p_{\text{accept}}^{ED}$, respectively. Their difference gives an estimate of the number of accepted shots with non trivial syndrome in the final decoding since
\begin{equation}
    p_{\text{accept}}^{EC}-p_{\text{accept}}^{ED}=\frac{N_{\text{post}}^{EC}}{N_{\text{tot}}}-\frac{N_{\text{post}}^{ED}}{N_{\text{tot}}}\:,
    \label{eq:dif_acceptance}
\end{equation}
Moreover, in Fig.~\ref{fig:difference_plot} we also show a plot of the difference of the weighted infidelities, which are defined as the infidelity times the acceptance rate
\begin{equation}
p_{\text{fail}}^{\alpha}\:p_{\text{accept}}^{\alpha}=\left(\frac{N^{\alpha}_{\text{fail}}}{N^{\alpha}_{\text{post}}}\right)\left(\frac{N^{\alpha}_{\text{post}}}{N_{\text{tot}}}\right)=\frac{N^{\alpha}_{\text{fail}}}{N_{\text{tot}}}\:,
\end{equation}
where $N_{\text{fail}}^{\alpha}$ indicates the number of shots with a logical failure and $\alpha=EC,ED$ (standing from error-correction and error-detection/postselection, respectively). The difference between the weighted infidelities gives an estimate of the number of logical failures with non-trivial syndrome in the final decoding 
\begin{equation}
p_{\text{fail}}^{EC}\:p_{\text{accept}}^{EC}-p_{\text{fail}}^{ED}p_{\text{accept}}^{ED}=\frac{N^{ED}_{\text{fail}}}{N_{\text{tot}}}-\frac{N^{EC}_{\text{fail}}}{N_{\text{tot}}}\:.
\label{eq:weigth_inf}
\end{equation}

A complementary check of the previous results, which is corroborated in Fig.~\ref{fig:difference_plot}, is that the difference of weighted infidelities (\ref{eq:weigth_inf}) is always smaller than the difference of acceptance rates (\ref{eq:dif_acceptance}).

It is worth pointing out that for $p \in [10^{-3},10^{-2}]$ the leading and sub-leading terms of the polynomial fits (\ref{eq:infidelity_fit}) have the same order of magnitude. Hence, even though the polynomial interpolates the data points in that regime, extrapolation for even lower probabilities $p \ll 10^{-3}$ might lead to unreliable results. A similar note should be addressed in relation to our reported result of $5.1 \times 10^{-7}$ in Table~\ref{tab:comparisson} for the infidelity of the post-selected state when $p=10^{-3}$. Notwithstanding, the extrapolation was done from a nearby error rate $p=5\times 10^{-3}$, and roughly twenty data points with different error rates were simulated in order to account for a robust interpolating fitting polynomial.

\section{Discussion}
\label{sec:discussion}

In this paper, we introduced a method for preparing a high-fidelity magic state using color codes. Our approach employs a judicious use of flag-based post-selection and a transversal gate (that was recently discovered in Ref.~\cite{sullivan2024}, which also appeared in Ref.~\cite{Heuben2024} during
the preparation of our manuscript.) An important question at this point is whether one can achieve an even higher-fidelity magic state by using a similar protocol, with a flag-based postselection approach on a larger 3D color code. There are several problems that need to be solved for such a study. First of all, we need to generalize our syndrome extraction circuit to the higher distance of the 3D codes. Secondly, a better simulation method must be developed. The infidelity of the magic state produced this way is expected to be multiple orders of magnitude lower than what is reported in this manuscript. Simulation of such a low error rate will be extremely challenging using a Monte Carlo approach. (For instance, Ref.~\cite{Gidney2024} reported weeks of performing a Monte Carlo simulation and it is not even clear if that will be enough for this study.) And, a naive density matrix simulation of the full circuit is likely out of reach for classical computers. Perhaps a tensor network calculation can be useful for this purpose. We leave these studies for future work.

Another important future research direction is an end-to-end analysis of the fidelity of the magic state. In order to use our high-fidelity magic state, one must convert the given state on the original 2D color code to a larger 2D lattice, employing postselection in this process. It was found in a recent study that there is a non-negligible difference between the logical error rate obtained before and after this process~\cite{Gidney2024}. The ratio between the two should depend on the details about the effective error model induced on the 2D color code patch which is influenced by the protocol. As such, it is not completely clear if the ratio reported in Ref.~\cite{Gidney2024} would remain the same whether their growing process is applied to our setup. Understanding this would require yet another challenging large-scale simulation.

Lastly, we note that our approach should be readily implementable in platforms that allow all-to-all connectivities such as the ion trap~\cite{Decross2024computational,Loschnauer2024scalable} and neutral atom-based quantum computers~\cite{Bluvstein2024}. Optimizing our protocol for these platforms and experimentally demonstrating a substantial improvement in the magic state infidelity will be an important experimental outcome that we anticipate to be achievable in the near term.

\section*{Acknowledgements}
LD was supported by funds from the University of California, Davis, and by the Dean’s Summer Graduate Support Award 2024  from the College of Letters and Science of the University of California, Davis. IK acknowledges support from NSF under award number QCIS-FF-2013562. We thank Craig Gidney for helpful discussions and Sascha Heußen for providing comments on the final manuscript.

\appendix
\section{Simulation Methods: further details}
\label{app:sim_details}

In this section we elaborate on the simulation methods.

To begin with, we give additional details about the derivation of (\ref{eq:fidelity_practical1}) and (\ref{eq:fidelity_practical2}) from (\ref{eq:non-convex}) and (\ref{eq:fidelity_formula}).  Notice that the final noisy state produced by the protocol $\bar{\rho}$ can be interpreted as the result of an effective linear map $\mathcal{E}$ acting on the noisy magic state $|\bar{T}\rangle \langle \bar{T}|$ prepared on the qRM code (\ref{eq:magic_qrm_noisy}), i.e., $\bar{\rho}=\mathcal{E}(|\bar{T}\rangle \langle \bar{T}|)$. Expressing the noisy magic state using (\ref{eq:non-convex}) leads to
\begin{equation}
\begin{aligned}
   \mathcal{E}( |\bar{T}{\rangle}{\langle} \bar{T}|) &=\frac{1}{2}I+\frac{\sqrt{2}}{4} \Big(\mathcal{E}( |\bar{+}{\rangle}{\langle} \bar{+}|)-\mathcal{E}( |\bar{-}{\rangle}{\langle} \bar{-}|)\Big)\\
    &+ \frac{\sqrt{2}}{4}\Big(\mathcal{E}( |\bar{Y}_+{\rangle}{\langle} \bar{Y}_+|) - \mathcal{E}( |\bar{Y}_-{\rangle}{\langle} \bar{Y}_-|)\Big).
\end{aligned}
\label{eq:non-convex_map}
\end{equation}

Therefore, the expected value $\langle\bar{X}\rangle_{\bar{\rho}}=\text{Tr}(\bar{X}\bar{\rho})$ appearing in (\ref{eq:fidelity_formula}) can be written as

\begin{equation}
    \langle \bar{X}\rangle_{\bar{\rho}}=\frac{\sqrt{2}}{4}\Big(\langle \bar{X}\rangle_{|\bar{+}\rangle}-\langle \bar{X}\rangle_{|\bar{-}\rangle}+\langle \bar{X}\rangle_{|\bar{Y}_+\rangle}-\langle \bar{X}\rangle_{|\bar{Y}_-\rangle}\Big)\:,
\label{eq:expected_Stab}
\end{equation}
and similarly for $\langle \bar{Y}\rangle_{\bar{\rho}}$. The lower index label indicates that the expected value is computed with respect to the respective noisy stabilizer state prepared on the qRM code subject to the linear map $\mathcal{E}$ (a process which has an associated circuit that we present in Fig.~\ref{fig:simulation_circuit}). Equations~(\ref{eq:fidelity_practical1}) and (\ref{eq:fidelity_practical2}) follow directly from (\ref{eq:expected_Stab}) and its analogous expression for $\langle \bar{Y}\rangle_{\bar{\rho}}$.

Finally, we discuss the importance of the ordering of the gates appearing in the Clifford error frame $C(P)$ (\ref{eq:CP}). In an earlier attempt, we implemented an incorrect ordering, which led to a higher logical error rate than the correct one. Specifically, although $C(\bar{X}) = \bar{S} \bar{X}$, we instead incorrectly implemented $\bar{X} \bar{S}$. (Note that in the second case a logical-$X$ operator propagates to a logical error of the final state.) This led us to manually update the Pauli error frame to the minimum-weight one, which is equivalent up to stabilizers and a logical-$X$ operator, before applying the Clifford error. Ultimately, we employed the correct ordering together with the minimum-weight update of the Pauli error frame, even though the latter has no effect on the outcome of the experiment. (We have confirmed the correctness by comparing the simulations with and without the Pauli frame update.) Thus, the result of our simulation is correct, and the method described in Sec.~\ref{sec:simulation} can be used as is.

\begin{figure}[b]
\centering
\[
\begin{array}{c}
\centering
\quad \quad\quad\quad\quad\quad\Qcircuit @C=0.8em @R=.8em  {\lstick{\text{qRM:}\:\:|\bar{+}\rangle_{\text{noiseless}}} &\gate{\bar{Q}} & \gate{C(P)}   &\ctrl{1}   & \measuretab{M_{\bar{X}}} & \control \cw &  \\
\lstick{\text{Steane:}\quad\quad\:\:\:\:|\bar{0}\rangle} &\qw & \qw &\targ & \qw & \gate{\bar{Z}} \cwx & \qw &
}
\end{array}
\]
\caption{Diagram of logical circuits utilized for the numerical simulations. $C(P)$ represents the Clifford error (\ref{eq:CP}) for a Pauli frame $P$ (\ref{eq:pauli_frame}), and $\bar{Q}\in \{I,\bar{Z},\bar{S},\bar{S}^{\dagger}\}$ represents the Pauli rotations which produces the stabilizer states $|\bar{+}\rangle$, $|\bar{-}\rangle=\bar{Z}|\bar{+}\rangle$, $|\bar{Y}_{+}\rangle=\bar{S}|\bar{+}\rangle$ and $|\bar{Y}_{-}\rangle=\bar{S}^{\dagger}|\bar{+}\rangle$. At the end of each circuit, after an ideal round of error correction (or post-selection), the expected value $\langle \bar{M}\rangle_{\bar{\rho}'_Q}$ with $\bar{M} \in \{\bar{X},\bar{Y}\}$ of the final noisy stabilizer state $\bar{\rho}'_Q$ is computed.} \label{fig:simulation_circuit} 
\end{figure}

\bibliography{bib}

\end{document}